\documentclass[
    ,a4paper
    ,aps,superscriptaddress
    ,pop
    ,oneside
    ,nobibnotes
    ,showpacs
    ,twocolumn
]{revtex4-1}

\usepackage{graphicx}
\graphicspath{{}{Graph/}{Graph/seminar/}}

\usepackage{amsmath,amssymb,bm}

\allowdisplaybreaks[1]

\usepackage{siunitx}
\usepackage{ifluatex,ifxetex}
\ifnum 0\ifxetex 1\fi\ifluatex 1\fi>0
    \usepackage{fontspec}
    \defaultfontfeatures{Ligatures={TeX},Renderer=Basic}

    \AtBeginDocument{\def\perp{\mathrel\bot}}

    \usepackage[math-style=ISO,bold-style=ISO]{unicode-math}
    \mathversion{cambria}
\else
    \usepackage[T2A,T1]{fontenc} 
    \usepackage[utf8]{inputenc} 
\fi

\usepackage[main=english,russian,showlanguages]{babel}

    \DeclareMathOperator{\e}{e}

    \DeclareMathOperator{\re}{Re}
    \DeclareMathOperator{\im}{Im}
    
    \DeclareMathOperator{\sign}{sign}

    \newcommand{\dif}[1][]{\mathop{}\!\mathrm{d}
        \if\relax\detokenize{#1}\relax
    \else
        ^{\mkern-1.mu#1}\mkern-2.5mu 
    \fi}
    \newcommand{\der}[2]{\frac{\dif{#1}}{\dif{#2}}}
    \newcommand{\tder}[2]{{\dif{#1}}/{\dif{#2}}}

    \usepackage[svgnames]{xcolor}
    \definecolor{darkblue}{cmyk}{1.00, 0.50, 0.00, 0.40}
\begin{document}

\def\perp{\mathrel\bot}

\title{Electron radiative recombination with a hydrogen-like ion}
\author{Igor A. Kotelnikov}
    \email{I.A.Kotelnikov@inp.nsk.su}
\author{Alexander I. Milstein}
\email{A.I.Milstein@inp.nsk.su}
    \affiliation{Budker Institute of Nuclear Physics, Novosibirsk, Russia}
    \affiliation{Novosibirsk State University, Novosibirsk, Russia}


\date{\today}

\begin{abstract}
We survey the results of a long-term study of the process of radiative recombination. A rigorous theory of  nonrelativistic electron radiative recombination   with a hydrogen-like ion is used to calculate the total cross section of the process, the effective radiation, the recombination rate coefficient, and the emission coefficient in a plasma with a Maxwellian electron distribution. The exact results are compared with the numerous known asymptotic and interpolation formulas. We propose interpolation formulas which ensure a uniform approximation of all mentioned quantities in a wide range of plasma temperatures.
\end{abstract}
\maketitle

\section{Introduction}\label{1}

The investigation  of radiative recombination (RR) has a long history. The process was the object for several theoretical papers starting already in the 1920s \cite{Kramers1923PM_46_836, Oppenheimer1927ZfPhysik_41_268, Oppenheimer1928PhysRev_31_349, Oppenheimer1929ZfPhysik_55_725, Stobbe1930AnnPhys_399_661, Wessel1930AnnPhys_5_611, Gaunt1930RSTA_229_163, MenzelPekeris1935MnNRAS_96_77}. Later, this process was investigated theoretically in numerous papers \cite{BetheSalpeter1957quantum, Kogan1958FizPlazUTS3_99(eng), LeePratt1976PhysRevA_14_990, HahnRule1977JPhysB_10_2689, KatkovStrahovenko1978SovPhysJETP_48_639, Kim+1983PhysRevA_27_2913, Milstein1989PhysLettA_136_52,  Milstein1990SovPhysJETP_70_982}. The main problem has been to find appropriate approximations to overcome essential difficulties in numerical calculations involved.

In 1923 Kramers \cite{Kramers1923PM_46_836} used semiclassical arguments to derive the cross section for RR of electron  been captured to highly excited states of a hydrogen atom. At that time the  level of quantum theory of radiation did not allow to solve the problem of radiative recombination consistently. Instead the idea of Bohr's correspondence principle was used to obtain an approximate solution.

Using Shr\"{o}dinger quantum mechanics, Oppengeimer suggested the solution of the problem in Refs.~\cite{Oppenheimer1927ZfPhysik_41_268, Oppenheimer1928PhysRev_31_349, Oppenheimer1929ZfPhysik_55_725}. Note that the asymptotic approximations of his formulas are in error.

Since the initial and final wave functions are known analytically in the Coulomb field of a bare ion, the cross section of radiative recombination in this case may in principle be calculated exactly. For capture to the lowest levels, this was done in the early work of Stobbe \cite{Stobbe1930AnnPhys_399_661}. As it was shown by Gaunt \cite{Gaunt1930RSTA_229_163},  the results of original work of Kramers  are valid in the region of large quantum numbers.

In the pioneering work of Menzel and Pekeris \cite{MenzelPekeris1935MnNRAS_96_77}, the process of radiation recombination was discussed in detail (their results were later corrected by Burgess \cite{Burgess1958NBRAS_118_477}) though at that time  a numerical analysis of the cross section of the process was a rather complicated problem, especially for  the total cross section. Much later Bethe and Salpeter  \cite{BetheSalpeter1957quantum} derived an approximate analytical formula for the cross section of radiative recombination, which in fact coincides with that of Kramers. In their work, advantage was taken of the fact that the oscillator strength crosses the continuum limit smoothly. Thus, knowledge of the bound-bound transition amplitudes was used to derive the cross section of RR. The Bethe-Salpeter formula provides satisfactory accuracy even for the ground and low excited states (see below).

Various other asymptotic formulas are used in the applications, as well as approximate formulas obtained by interpolating the results that are valid for large and small energies of the incident electron (see \cite{Kogan1958FizPlazUTS3_99(eng)}). An important step has been made by Katkov and Strakhovenko in their paper \cite{KatkovStrahovenko1978SovPhysJETP_48_639}, where a  simple expression was obtained for the cross section $\sigma_{\text{rr}}^{(n)}$ of a free electron radiative capture to a level with an arbitrary principle value $n$ of a hydrogen-like ion.

Finally, in two papers of Milstein \cite{Milstein1989PhysLettA_136_52, Milstein1990SovPhysJETP_70_982} the relatively simple expressions for the total cross section $\sigma_{\text{rr}}$ and the total effective radiation $\varkappa_{\text{rr}}$ of the process were derived, where
    \begin{equation}
    \label{1:01}
    \sigma_{\text{rr}}=\sum_{n}\sigma_{\text{rr}}^{(n)}\,,
    \quad \varkappa_{\text{rr}}=\sum_{n}\hbar\omega_{n}\sigma_{\text{rr}}^{(n)}\,,
\end{equation}
 $\omega_{n}$ is a photon frequency, $\hbar\omega_{n}=\varepsilon+J_{Z}/n^{2}$, $\varepsilon=m_ev^2/2$, $J_{Z}=Z^2e^4m_e/(2\hbar^2)$, $v$ is the electron velocity, $Z$ is the ion charge number, $e$ and $m_e$ are the electron charge and mass, respectively. The results were obtained  using a dipole approximation and the analytical  properties of the electron Green's function in a Coulomb field.

Although the RR process has been treated theoretically for many years, the first successful direct measurement of the process was not done until 1990s, when Anderson with coauthors succeeded in measuring the recombination rate coefficient $k_\text{rr}$ for electron radiative recombination on bare carbon ions \cite{Andersen+1990PhysRevLett_64_729},
    \begin{equation}
    \label{1:02}
    k_\text{rr}
    =
    \langle v\sigma_{\text{rr}}\rangle
    =
    \int v \sigma_\text{rr} f_{e}(v) \dif^{3}{v}\,
    ,
    \end{equation}
 where $f_{e}(v)$ is the Maxwellian electron distribution function.
They also measured $ k_\text{rr}$  for hydrogen and a few hydrogen-like ions and compared data  with a calculation based on the theory of Stobbe, Bethe and Salpeter \cite{Andersen+1990JPhysB_23_3167, Andersen+1990PhysRevA_42_1184}. Direct observations of electron-ion recombination in colliding-beam experiment was reported in Ref.~\cite{Spies+1992PhysRevLett_69_2768}.

The study of radiative recombination of elementary particles in the Coulomb field of a nucleus or an ion is quite important in several areas  of physics: particle accelerators, plasma physics, astrophysics, antimatter production, and laser-induced recombination.
For instance, in the  particle accelerator physics the electron cooling of proton beams should be mentioned first \cite{BudkerSkrinsky1978UFN_124_561(eng), ColeMills1981AnnuRevNPS_31_295, Bell+1981PartAccel_12_49, Poth1990Nature_345_399, Poth1990PhysRep_196_135, ParkhomchukSkrinsky2000UFN_170_473(eng)}. The cooling mechanism of protons by electrons consists of a redistribution of energy during  collision of the two gases.

Laser-induced recombination has been investigated both theoretically and experimentally with  special emphasis on an enhancement of the recombination cross sections to moderately-high excited states \cite{Neumann+1983ZPhysAtomNuclei_313_253, Schramm+1991PhysRevLett_67_22, Yousif+1991PhysRevLett_67_26, Badnell+1992PhysRevA_45_2820, Tanabe+1992PhysRevA_45_276, Vardi+1997JChemPhys_107_6166, Amoretti+2006PhysRevLett_97_213401}.

Laser-stimulated electron-proton recombination for the conditions encountered in ion storage rings was first examined in Ref.~\cite{Neumann+1983ZPhysAtomNuclei_313_253}, and the first observations of laser-induced recombination obtained with merged beams of protons and electrons were reported in Refs.~\cite{Schramm+1991PhysRevLett_67_22, Yousif+1991PhysRevLett_67_26}. The laser-mediated electron-proton recombination rate coefficient for a steady state situation is given in \cite{Wolf1993HyperfineInteration_76_189}.

Lately, new attention has been paid to the RR process, since the process has been suggested as a method to produce antimatter  \cite{BudkerSkrinsky1978UFN_124_561(eng), Poth1988PhysScripta_38_806}.  The radiative recombination of a positron and an antiproton will produce antihydrogen. In order to increase the rate of antihydrogen production it has been proposed to stimulate the recombination process by a laser \cite{Neumann+1983ZPhysAtomNuclei_313_253}.

In plasma physics, when an electron emits a photon in collision with ions, the electron energy decreases and free-free or free-bound transitions can be observed. In the latter case an electron can be captured into any energy level of recombined atom. The study of recombination radiation is important, for instance, in determining the rate at which positive ions recapture electrons, although it is usually believed that at higher temperatures recombination radiation is negligible compared to bremsstrahlung
\cite{Spitzer1962Physics, Tucker1978radiation, McDaniel1964collision, MichelisMattioli1981NF_21_677, KoganLicitsa1983VINITI_4_194(eng), MichelisMattioli1984RepProgrPhys_47_1233}.

Plasma physics operates, in addition to the recombination rate coefficient $k_\text{rr}$, with  the  emission coefficient $q_{\text{rr}}$,
\begin{equation}
    q_\text{rr}
    =
    \langle v\varkappa_{\text{rr}}\rangle=\int
    v \varkappa_\text{rr}f_{e}(v) \dif^{3}{v}\,.
\end{equation}
 The most well-known calculations of these quantities were made by Seaton \cite{Seaton1959MNRAS_119_81}. Using the first three terms in the asymptotic expansion of the Gaunt factor, he calculated  the recombination  rate coefficient and the mean kinetic energy of the recombining electrons. Seaton has shown that if   one  put the Gaunt factor equal to unity, then the obtained  relatively simple expressions will have   errors not more than  20 percent. A considerable improvement was obtained by using the asymptotic expansion of the Gaunt factor, as derived by Menzel and Pekeris \cite{MenzelPekeris1935MnNRAS_96_77} and corrected by Burgess \cite{Burgess1958NBRAS_118_477}. From the expansion of the Gaunt factor, Seaton obtained asymptotic expansions which enable the rate coefficients to be calculated with errors which do not exceed $2\%$ for temperatures of order $10^{4}\,\text{K}$ or less but which may be greater for temperatures of order $10^{6}\,\text{K}$. He presented a systematic tabulation of the various functions which occur in these asymptotic expansions and derived a parametrization for the rate coefficient. It is this parametrization that is cited by the well-known reference book of formulas for the physics of plasma \cite{NRL2016}.

Further studies of RR cross sections, effective radiation, rate and emission coefficients following Seaton's work are given in
Refs.~\cite{%
    HummerSeaton1963MNRAS_125_437,
    BrownMathews1970ApJ_160_939,
    Tarter1971ApJ_168_313, 
    Tarter1973ApJ_181_607, 
    AldrovandiPequignot1973AA_25_137, 
    Vainstein+1973CrossSections(eng), AldrovandiPequignot1974RevBrasFis_4_491, 
    Goeler+1975NF_15_301,
    Breton+1976NF_16_891,
    AldrovandiPequignot1976AA_47_321, 
    Breton1978JQSRT_19_367,
    Bottcher1979NATO_53_267,
    Barfield1980JPhysB_13_931,
    Ferland1980PASP_92_596,
    Martin1988ApJS_66_125, 
    Pequignot+1991AA_251_680, 
    Erdas+1993PhysRevA_48_452,
    ErdasQuarati1994ZPhysD_31_161,
    Hahn1997RepProgPhys_60_691%
}, where the coefficients for hydrogen-like ions were calculated in Refs.~\cite{%
    Vainstein+1973CrossSections(eng),
    Barfield1980JPhysB_13_931,
    Ferland1980PASP_92_596,
    Martin1988ApJS_66_125,
    Pequignot+1991AA_251_680,
    Erdas+1993PhysRevA_48_452,
    ErdasQuarati1994ZPhysD_31_161,
    Hahn1997RepProgPhys_60_691%
}. In particular, accurate piecewise-continuous parametrization for the total cross section is proposed by Erdas and coauthors in Ref.~\cite{Erdas+1993PhysRevA_48_452} (whereas their parametrization of the total effective radiation coefficient seems erroneous).
%
%
Review paper \cite{Hahn1997RepProgPhys_60_691} discusses also measurements of the recombination rate coefficient although for the temperatures below $1\,\text{eV}$.
Apart of Seaton's work, the recombination rate coefficient for hydrogen and some other ions in the form of four-parameter fits are presented in Ref.~\cite{Pequignot+1991AA_251_680} for relatively narrow range of temperatures $0.004\,\text{eV}<T<2\,\text{eV}$. A more complicated formula for the recombination rate coefficient $k_\text{rr}$, containing the summation of an infinite series, was used in
Refs.~\cite{Goeler+1975NF_15_301, Breton+1976NF_16_891, Breton1978JQSRT_19_367, Kim+1983PhysRevA_27_2913}.
Tabulations of fits to the recombination rate coefficient $k_\text{rr}$ and/or emission coefficient $q_\text{rr}$
have been done by Tarter \cite{Tarter1971ApJ_168_313, Tarter1973ApJ_181_607}, Aldrovandi \&
Pequignot \cite{AldrovandiPequignot1973AA_25_137, AldrovandiPequignot1974RevBrasFis_4_491, AldrovandiPequignot1976AA_47_321}, Martin \cite{Martin1988ApJS_66_125}, Erdas \& Quarati \cite{ErdasQuarati1994ZPhysD_31_161}.

All known parametrizations of $\sigma_{\text{rr}}$, $ \varkappa_{\text{rr}}$, $k_\text{rr}$, and
$q_\text{rr}$ were proposed at the time when Milstein's formulas
\cite{Milstein1989PhysLettA_136_52, Milstein1990SovPhysJETP_70_982}
 had not yet gained wide popularity. In the present paper we fill this gap by averaging Milstein's formulas over the Maxwellian distribution of the electrons. The results of averaging are presented in the form of uniform parametrizations. In the entire temperature range of practical interest, from the smallest to the largest temperature values, these parametrizations ensure the accuracy of approximation at a level not exceeding several percents. Comparing the results of our calculation with the known parametrizations and tabulations, we  refine their range of applicability.

The above-mentioned publications date back to the late 1990s. More recent studies of radiative recombination focus primarily on processes involving partially ionized atoms heavier than hydrogen. We intentionally leave this vast area of research beyond the scope of this article.

Below we adhere to the following outline. In Section \ref{2} we present the main expressions for the cross section of electron radiative recombination with hydrogen-like ion using in our calculations. In Section \ref{3} we do the same for the effective radiation. In Sections~\ref{4} and \ref{5} we compute the rate and emission coefficients of the radiative recombination. Finally, in Section \ref{9} we summarize our results.

\selectlanguage{english}

\section{Cross section of radiative recombination}\label{2}

To begin with, we recall the results on the cross section of radiative recombination known from the literature.

\subsection{Recombination to the ground state, $n=1$}\label{2.1}

If the cross section  $\sigma_\text{pi}^{(n)} $ of the photoionization of a quantum level with the principal quantum number $n$ is known, then the radiative recombination cross section $\sigma_\text{rr}^{(n)} $ to this level can be found using the principle of detailed balance, according to which
    \begin{equation}
    \label{2.1:01}
    \sigma_\text{rr}^{(n)}
    =
    2\frac{(\hbar\omega_{n}/c)^2}{(m_{e}v)^2}\,
    \sigma_\text{pi}^{(n)}
    ,
    \end{equation}
where $c$ is the speed of light.  A quantum theory of photoionization was developed by Stobbe in 1930 \cite{Stobbe1930AnnPhys_399_661}. For reference, we remind some results of this theory \cite[\S56, Eqs.~(56.13) and (56.14)]{LLIV1982(eng)}. The photoionization cross section of the ground level ($n=1$) of the hydrogen-like ion at the ionization threshold at $\hbar\omega = J_{Z}$ is equal to
    \begin{equation}
    \label{2.1:02}
    \sigma_\text{pi}^{(1)}=
    \frac{2^9\pi^2}{3\e^4}\frac{\alpha a_\text{B}^2}{Z^2}
    ,
    \end{equation}
where  $a_{\text{B}} = \hbar^{2}/m_{e}e^{2}$ is the Bohr radius, $\e=2{,}71{\ldots}$ (do not confuse with elementary charge $e$).
For $\hbar\omega\gg J_{Z}$, the photoionization cross section of the ground level,
    \begin{equation}
    \label{2.1:03}
    \sigma_{\text{pi}}^{(1)}
    =
    \frac{2^8\pi}{3} \frac{\alpha a_\text{B}^2}{Z^2}
    \left(\frac{J_{Z}}{\hbar\omega}\right)^{7/2},
    \end{equation}
rapidly decreases with increasing photon energy $\hbar\omega$.
Combining Eqs.~\eqref{2.1:02} and~\eqref{2.1:03} with Eq.~\eqref{2.1:01}, we find the radiative recombination cross section to the ground level of the hydrogen-like ion,
    \begin{equation}
    \label{2.1:04}
    \sigma_\text{rr}^{(1)}
    =
    \frac{2^8\pi^2}{3\e^4}\,\alpha^3\,
    a_\text{B}^2\,
    \frac{J_{Z}}{\varepsilon}
    \end{equation}
at $\varepsilon\ll J_{Z}$, and
    \begin{equation}
    \label{2.1:05}
    \sigma_\text{rr}^{(1)}
    =
    \frac{2^7\pi}{3}\,\alpha^3
    a_\text{B}^2
    \left(\frac{J_{Z}}{\varepsilon}\right)^{5/2}
    .
    \end{equation}
at $\varepsilon\gg J_{Z}$.
The exact solution of the problem of radiative recombination to the ground state of a hydrogen-like ion was found by Stobbe \cite{Stobbe1930AnnPhys_399_661} (see also \cite[\S56]{LLIV1982(eng)}):
    \begin{equation}
    \label{2.1:06}
    \sigma_{\text{rr}}^{(1)}
    =
    \frac{
        2^{8} \pi ^2
    }{3}
    \frac{ \eta ^6
        \e^{
            -4 \eta\arctg\left({1}/{\eta }\right)
        }
    }{
        \left(1-\e^{-2 \pi\eta }\right)
        \left(\eta ^2+1\right)^2
    }\,
    \alpha ^3 a_{\text{B}}^2
    ,
    \end{equation}
where $\eta=Ze^{2}/\hbar v=\sqrt{J_{Z}/\varepsilon }$.
This expression agrees with the approximate formulas \eqref{2.1:04} and \eqref{2.1:05} in the limiting cases $\eta\gg1$ and $\eta\ll1$, respectively.

\subsection{Recombination to excited states, $n>1$}\label{2.2}

In the limit $\varepsilon \ll J_{Z}$ ($\eta\gg1$), there is a semiclassical formula of Kramers \cite[p.~861, Eq.~(65)]{Kramers1923PM_46_836}
    \begin{equation}
    \label{2.2:01}
    \sigma_\text{Kramers}^{(n)}
    =
    \frac{32\pi}{3\sqrt{3}}\,
    \frac{\eta^{4} \alpha ^{3}a_{\text{B}}^{2}}{n\left(\eta^{2}+n^{2}\right)}
    .
    \end{equation}
It was anew obtained as a result of simplifying formulas of quantum theory in the monograph of Bethe and Salpeter \cite[Eq.~(75.7)]{BetheSalpeter1957quantum}. A heuristic derivation of the Kramers formula is given by Kogan and Lisitsa in Ref.~\cite[p.~210, Eq.~(19)]{KoganLicitsa1983VINITI_4_194(eng)}. In the monographs of Raizer \cite{Raizer1987(eng), Raizer1992(eng), Raizer2009(eng)} a similar derivation is called ``adventurous''.

\begin{figure}[tb]
  \centering
  \includegraphics[width=\linewidth]{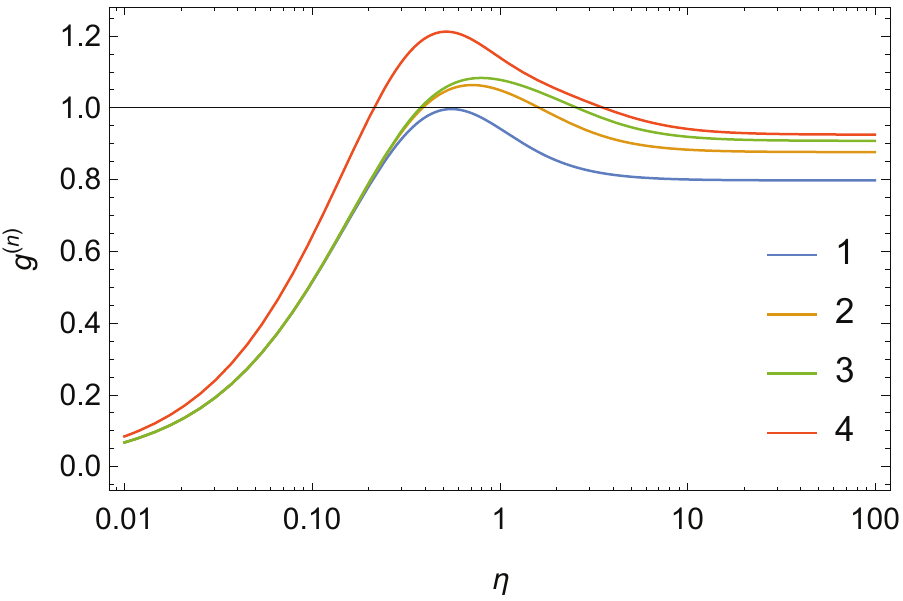}
  \caption{
    Gaunt factor $g^{(n)} = \sigma_{\text{rr}}^{(n)}/\sigma_{\text{Kramers}}^{(n)}$
    as a function of the parameter
    $\eta = \sqrt{J_{Z}/\varepsilon} $ for $n=1,2,3,4$.
  }\label{fig:Sigma(k)ToSigmaKramers(k)}
\end{figure}
Exact formulas in $\eta$ were found by Katkov and Strakhovenko in Ref.~\cite{KatkovStrahovenko1978SovPhysJETP_48_639}. After summing over the orbital quantum number $l$, they derived  the radiative recombination cross section to an arbitrary state with the principal quantum number $n$,
\begin{equation}
    \label{2.2:02}
    \sigma_{\text{rr}}^{(n)}
    =
    \frac{
        2^{8} \pi ^2
    }{3}
    \frac{
        \eta^6
        \e^{
            -4 \eta\arctg\left({1}/{\eta }\right)
        }
    }{
        \left(1-\e^{-2 \pi\eta }\right)
        \left(\eta ^2+n^{2}\right)^2
    }\,
    S_{n}(\eta)
    \,\alpha ^3 a_{\text{B}}^2
    ,
    \end{equation}
where the coefficient $S_{n}(\eta) $ for several lower levels has the form
    \begin{equation}
    \label{2.2:03}
    \begin{gathered}
    S_{1} = 1
    ,\\
    S_{2} = 2 + \frac{3}{x_{2}} + \frac{1}{x_{2}^{2}}
    ,\\
    S_{3} = 3 + \frac{14}{x_{3}} + \frac{19}{x_{3}^{2}}
        + \frac{8}{x_{3}^{3}} + \frac{1}{x_{3}^{4}}
    ,\\
    S_{4} = 4 + \frac{38}{x_{4}} + \frac{346}{3x_{4}^{2}}
        + \frac{409}{3x_{4}^{3}} + \frac{622}{9x_{4}^{4}}
        + \frac{43}{9x_{4}^{5}} + \frac{1}{x_{4}^{6}}
    ,
    \end{gathered}
    \end{equation}
where $x_{n} = (n^{2}+\eta^{2})/4\eta^{2}$. Figure~\ref{fig:Sigma(k)ToSigmaKramers(k)} shows the dependence of the Gaunt factor
    \begin{equation}
    \label{2.2:04}
    g^{(n)}(\eta) = \sigma_{\text{rr}}^{(n)}(\eta)/\sigma_{\text{Kramers}}^{(n)}(\eta)
    \end{equation}
on $\eta$ for a few lowest levels. It  clearly demonstrates that the Kramers formula gives a more or less correct result only for $\eta\gg 1 $. Although it was derived under the assumption $n\gg1$, for $n=1$ and $\eta\gg 1 $ it gives the result which is less than that of Stobbe \eqref {2.1:04} only by $20\%$,
    \begin{equation}
    \label{2.2:05}
    g^{(1)}(\infty)
    =
    \frac{8\sqrt{3}\pi}{\e^{4}} = \num{0.797301}
    .
    \end{equation}
%
For other values of $n$ in the limit $\eta\to\infty$ the error is $12\% $ for $n=2$, $9\%$ for $n=3 $, and $7.5\%$ for $n=4$:
    \begin{gather}
    \begin{gathered}
    g^{(2)}(\infty) = \frac{480\sqrt{3}\pi}{\e^{8}}
        = \num{0.876185137748039}
    ,\\
    g^{(3)}(\infty) = \frac{27144\sqrt{3}\pi}{\e^{12}}
        = \num{0.9075082124499262}
    ,\\
    g^{(4)}(\infty) = \frac{13591712\pi}{3\sqrt{3}\e^{16}}
        = \num{0.9247629967974488}
    .
    \end{gathered}
    \end{gather}
In other words, the Kramers formula can even be used for $n=1$, if the error of the order of $20\%$ is not burdensome.

Below we enlist some papers where Kramers' formula \eqref{2.2:01} was used or cited with a mention of his name:
    \cite[Chapter~5, \S4, Eq.~(5.27)]{ZeldovichRaizer1966AcademicPress};
    \cite[Chapter~9, \S3, Eq.~(9.5)]{Raizer1987(eng)};
    \cite[Chapter~6, \S3, Eq.~(6.5)]{Raizer1992(eng)};
    \cite[Chapter~8.3.1, p.~245]{Raizer2009(eng)};
    \cite[p.~210, Eq.~(19)]{KoganLicitsa1983VINITI_4_194(eng)};
    \cite[Eq.~(16)]{Milstein1990SovPhysJETP_70_982};
    \cite[Eq.~(3.1)]{Hahn1997RepProgPhys_60_691}.
The formula \eqref {2.2:01} is cited by M.~Bell \& J.~Bell in Ref.~\cite{Bell+1981PartAccel_12_49} with reference to Spitzer's book \cite{Spitzer1962Physics}, and neither Bell \& Bell nor Spitzer mentioned the name of Kramers. Andersen attributed the formula \eqref {2.2:01} to Bethe and Salpeter \cite[Eq.~(3)]{Andersen+1990PhysRevA_42_1184}.

\subsection{Total cross section of radiative recombination}\label{2.3}

The total cross section $ \sigma _{\text{rr}}$ of radiative recombination is defined as a sum of the recombination cross sections to individual levels.
In an idealized model, the number of levels is infinite, so that the summation in the formula \eqref{1:01} goes from $n=1$ to $n=\infty$. Note that in a real plasma the number of levels is finite.

A contemporary quantum theory of radiative recombination for a hydrogen-like ion is described in Milstein's papers \cite{Milstein1989PhysLettA_136_52, Milstein1990SovPhysJETP_70_982}, where summation over $n$ is performed  and the following expression for the total  cross section is obtained \cite[Eq.~(12)]{Milstein1990SovPhysJETP_70_982}:
\begin{widetext}
    \begin{equation}\label{cs0}
    \sigma_{\text{rr}}
    =
    -\frac{16}{3} \pi ^2 \alpha ^3 a_{\text{B}}^2 \eta ^2
    \left[
        \int_{0}^{\infty}
        \frac{
            \sign(\varepsilon'-\varepsilon)
        }{
            \varepsilon -\varepsilon'
        }\,
        \frac{
            \sinh(\pi \eta-\pi\eta')
        }{
            2 \sinh(\pi \eta) \sinh(\pi \eta')
        }
        \left(
            \xi \der{}{\xi}|F(\xi)|^{2}
        \right)
        \dif{\varepsilon'}
        +
        \coth(\pi\eta)
        -
        \frac{1}{\pi\eta}
    \right]
    .
    \end{equation}
Here $\eta =\sqrt{J_{Z}/\varepsilon}$, $\eta' =\sqrt{J_{Z}/\varepsilon'}$, $\xi=-4\eta\eta'/(\eta-\eta')^{2}$, and
    \(
    F(\xi)
    =
    {_{2\!}F_{1}}(i\eta, i\eta', 1; \xi)
    \)
is expressed in terms of the hypergeometric function ${_{2\!}F_{1}}$. Passing in this formula to integration over the variable $ \eta '$, we have
    \begin{equation}
    \label{2.3:03}
    \sigma_{\text{rr}}
    =
    \frac{16}{3} \pi ^2 \alpha ^3 a_{\text{B}}^2 \eta ^2
    \left[
        \int_{0}^{\infty}
                \frac{\eta ^2
            \sinh(\pi \eta-\pi\eta')
        }{\eta'\,|\eta^{2}-\eta'^{2}|
            \sinh(\pi \eta) \sinh(\pi \eta')
       }
    \left(
            \xi \der{}{\xi}|F(\xi)|^{2}
        \right)
        \dif{\eta'}
        -
        \coth(\pi\eta)
        +
        \frac{1}{\pi\eta}
    \right]
    .
    \end{equation}
\end{widetext}

As stated in Ref.~\cite{Milstein1990SovPhysJETP_70_982}, in the limit $ \eta \ll1 $ the result of integration coincides with the well-known formula
    \begin{equation}
    \label{2.3:04}
    \sigma_{\text{rr}}
    =
    \frac{128\pi\zeta(3)}{3}\,\eta^{5}\, \alpha ^3 a_{\text{B}}^2
    ,
    \end{equation}
where $ \zeta(x) = \sum_{k = 1}^{\infty} 1/k^{x}$ is the Riemann zeta function.
Using the fitting method  at $ \eta < 0.05$, we find a more accurate interpolation
    \begin{equation}
    \label{2.3:04a}
    \frac{\sigma_{\text{rr}}}{\eta^{5}\alpha ^3 a_{\text{B}}^2}
    =
    \frac{128\pi\zeta(3)}{3}
    \left[
        1
        -\num{3.1396}\,\eta
        +\num{5.14194}\,\eta^2
        -\num{0.169201}\,\eta^3
    \right]
    \end{equation}
for small $\eta$.

For the case $ \eta \gg 1$ Eq.~\eqref{2.3:03} is reduced  to the asymptotic expression given in Ref.~\cite{Milstein1990SovPhysJETP_70_982}
    \begin{equation}
    \label{2.3:05}
    \sigma_{\text{rr}}
    =
    \frac{32\pi}{3\sqrt{3}}\, \alpha ^3 a_{\text{B}}^2\,
    \eta^{2}\ln(\eta)
    .
    \end{equation}
The same result can be obtained from the Kramers formula
    \begin{equation}
    \label{2.3:06}
    \sigma_{\text{Kramers}}
    =
    \frac{32\pi}{3\sqrt{3}}\, \alpha ^3 a_{\text{B}}^2
    \sum_{n=1}^{\infty }\frac{\eta^{4}}{n\left(\eta^{2}+n^{2}
    \right)}
    ,
    \end{equation}
which is valid for $ \eta \gg 1 $. In this formula the main contribution to the sum  is given by large $n\sim\eta$. Replacing the summation over $n$ by integration from $1$ to $\infty$, we arrive at the formula
    \begin{equation}
    \label{2.3:07}
    \sigma_{\text{rr}}
    =
    \frac{16\pi}{3\sqrt{3}}
    \,
    \eta ^{2}\ln(1+\eta^{2})\,
    \alpha^3 a_{\text{B}}^2
    ,
    \end{equation}
that almost coincides with \eqref{2.3:05}. The latter formula is cited in Ref.~\cite[page~211, Eq.~(23)]{KoganLicitsa1983VINITI_4_194(eng)} and not only there. The sum in Eq.~\eqref{2.3:06} can also be calculated without going over to integration. This is done by Bell \& Bell in Ref.~\cite{Bell+1981PartAccel_12_49} (see Eq.~(9) there); the result is expressed in terms of the digamma function $\psi(z) = \Gamma'(z)/\Gamma(z)$ and has the form
    \begin{equation}
    \label{2.3:08}
    \sigma_{\text{Kramers}}
    =
    \frac{
        16 \pi
    }{3 \sqrt{3}}
    \,\eta^2
    \left[
        \psi(1+i\eta)
        +
        \psi(1-i\eta)
        +
        2\gamma
    \right]
    \alpha ^3 a_{\text{B}}^2
    ,
    \end{equation}
where $\gamma\approx\num{0.577216}$ is the Euler constant.
%
%
In Ref.~\cite{Bell+1981PartAccel_12_49} cited above other works of 1939-1969, Refs.~
\cite{Bates+1939RSPA_170_322, Burgess1958NBRAS_118_477, Seaton1959MNRAS_119_81, Massey+1969Oxford} concerning calculations of the total recombination cross section, were discussed. Authors of Ref.~\cite{Bell+1981PartAccel_12_49} proposed the approximate formula
    \begin{equation}
    \label{2.3:07a}
    \sigma_{\text{rr}}
    =
    \frac{32\pi}{3\sqrt{3}}\,
    \alpha^3 a_{\text{B}}^2
    \left[
        \ln(\eta)
        +
        \num{0.1402}
        +
        \num{0.525}\,
        \eta^{-2/3}
    \right]
    \end{equation}
for the case $\eta\gg1$. Our calculations has shown that it is valid with the accuracy better than $3.3\%$ for $\eta\ge10$  and better than $1\%$ for $\eta\ge50$.

Comparing the result of integration in the formula \eqref{2.3:03} with the Kramers formula \eqref{2.3:08}, we found that in the interval $ 50 < \eta <250 $ the formula
    \begin{equation}
    \label{2.3:09}
    \sigma _{\text{rr}}
    =
    \num{0.924841}\,\sigma_{\text{Kramers}}
    \end{equation}
approximates exact solution \eqref{2.3:03} with accuracy better than $1\%$, and the exact value is only $3.2\% $ less than the approximation \eqref{2.3:09} at $ \eta = 10$.

\begin{figure}
  \centering
  \includegraphics[width=\linewidth]{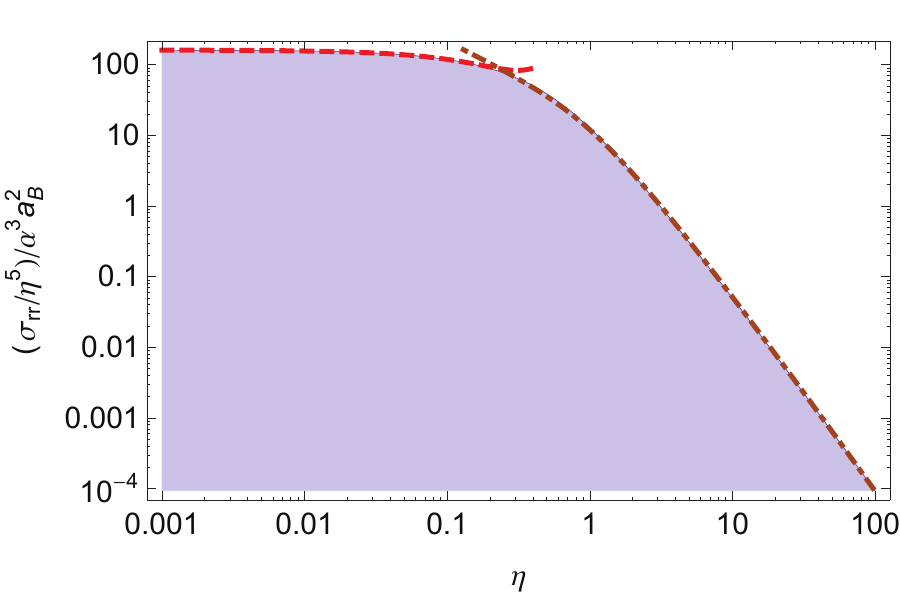}
  \caption{
       The ratio $\sigma_{\text{rr}} / \eta^{5}\alpha ^3 a_{\text{B}}^2$ (the edge of the shaded region), parametrization Eq.~\eqref{2.3:04a} for $\eta\ll1$ (dashed line) and Eq.~\eqref{2.3:09} for $\eta\gg 1$ (dash-dotted line) as a function of $\eta$.  The interpolations intersect at $\eta = \num{0.234146}$ where their join value exceeds the exact value by $10\%$.
  }\label{fig:SigmaTotalByEta5}
\end{figure}

Note that the asymptotics \eqref {2.3:04a} and \eqref{2.3:09} are joined at the point $\eta = \num{0.234093}$, where the value calculated by the exact formula \eqref{2.3:03} is $10\%$ less. Figure \ref{fig:SigmaTotalByEta5} shows the exact ratio $\sigma_{\text{rr}} / \eta^{5}\alpha ^3 a_{\text{B}}^2 $, together with its  asymptotics \eqref{2.3:04a} and \eqref{2.3:09}, as a function of $\eta$. This ratio  determines  the reaction rate of the radiative recombination (see Section~\ref{4}).

\begin{figure}[t]
  \centering
  \includegraphics[width=\linewidth]{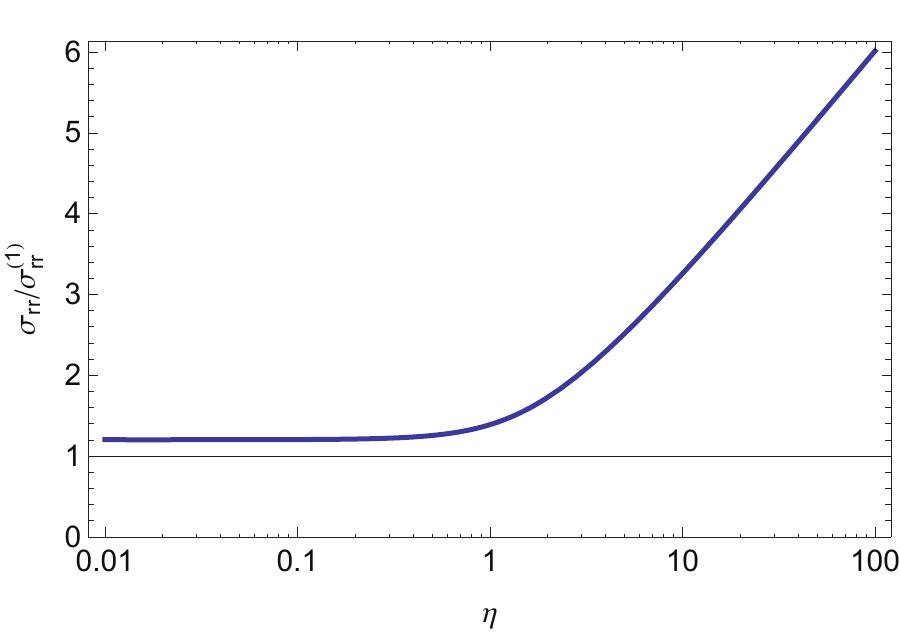}
  \caption{
    The  ratio $\sigma_{\text{rr}}/\sigma_{\text{rr}}^{(1)}$ as a function of $\eta$.}
  \label{fig:SigmaTotalToSigma1}
\end{figure}

Figure~\ref{fig:SigmaTotalToSigma1} shows the  ratio $\sigma_{\text{rr}}/\sigma_{\text{rr}}^{(1)}$ as a function of $\eta$. It is clearly seen  that for $\eta\gg 1 $ the main contribution to the total cross section is given  by  recombination to the numerous excited states of the recombined atom.

Concluding this section, we point out that an accurate  piecewise-continuous parametrization of the total cross section for radiative recombination was proposed by Erdas et al.\ in Ref.~\cite{Erdas+1993PhysRevA_48_452},
    \begin{widetext}
    \begin{equation}
    \sigma_{\text{rr}} = \sigma_{\text{rr}}^{(1)}
    \begin{cases}
      \num{1.2020569}
        , & \mbox{if } \eta < 10^{-1/2} \\
      \num{0.480383} + \num{1.1998079933748376}\ln\eta
        , & \mbox{if } \eta > 10 \\
      \num{1.39636} + \num{0.36605552713972145} \ln\eta + \num{0.19569368901100592} \ln^{2}\eta
        , & \mbox{otherwise}.
    \end{cases}
    \end{equation}
According to our calculations, the accuracy of this parametrization in the interval $\eta < 100$ is better than $1.5\%$ but the error approaches $5\%$ as $\eta\to\infty$. We found the uniform parametrization
    \begin{equation}
    \sigma_{\text{rr}}
    =
    \sigma_{\text{rr}}^{(1)}\,
    \frac{
         \num{1.20206}
        +\num{0.57815} \ln\left(\eta ^2+1\right)
        +\num{0.214805} \ln^2\left(\eta^2+1\right)
    }{
        1 + \num{0.342529} \ln \left(\eta^2+1\right)
    }
    .
    \end{equation}
The error is less than one percent.
    \end{widetext}

\section{Effective radiation}\label{3}
Let us now turn to the discussion of effective radiation $\varkappa_{\text{rr}}$, see Eq.~\eqref{1:01}, which also is an important characteristic of the process of radiative recombination;
it determines the emission coefficient (see Section~\ref{5}).
Using the analytical properties of the Green's function of electron in a Coulomb field,  Milstein in his  paper \cite{Milstein1990SovPhysJETP_70_982}
has obtained the following result:
\begin{widetext}
    \begin{multline}
    \label{4:03}
    \varkappa _{\text{rr}}
    =
    -\frac{16}{3} \pi^2 \alpha ^3 a_{\text{B}}^2 J_{Z}
    \eta^2
    \left\{
        \int_{0}^{\infty}
        \left[
            \frac{
                \left|\sinh(\pi\eta - \pi\eta')\right|
            }{
                2\sinh(\pi\eta)\,\sinh(\pi\eta')
            }
            \left(
                \xi \der{}{\xi}|F(\xi)|^{2}
            \right)
            -
            \frac{4\eta'}{\pi}
            +
            \frac{4\eta^{2}\eta'^{2}}{(\eta+\eta')^{2}}\coth(\pi\eta)
        \right]
        \frac{2\dif{\eta'}}{\eta'^{3}}
    \right.
    \\
    \left.
        +
        8\left[
            \left(
                2-\ln(4)
                +\psi(1)
                -\re\psi(1+i\eta)
            \right)
            \coth(\pi\eta)
            +
            \frac{1}{\pi\eta}
            +
            \frac{1}{\pi}
            \im\psi'(1+i\eta)
        \right]
    \right\}
    .
    \end{multline}
where $\psi(x)=\Gamma'(x)/\Gamma(x)$, $\psi'(x)=\tder{\psi(x)}{x}$, and other notation are given after Eq.~\eqref{cs0}.
\end{widetext}
In the limiting case $\eta\ll1$ this yields
    \begin{equation}
    \label{4:06}
    \varkappa_{\text{rr}}
    =
    \frac{128\pi}{3}\,\zeta(3)\,
    \alpha ^{3}a_{\text{B}}^{2}\,J_{Z}\,\eta^{3}
    .
    \end{equation}
The asymptotics \eqref {4:06} corresponds to the fact that $\hbar\omega\approx\varepsilon$ for $\eta\ll1$, and the formula \eqref{2.3:04} can be used.

For $ \eta \gg 1 $, numerical integration gives the following asymptotic behavior:
    \begin{equation}
    \label{4:07}
    \varkappa_{\text{rr}}
    =
    \frac{128\pi}{3}\,\zeta(3)\,\num{0.117443}\,
    \alpha ^{3}a_{\text{B}}^{2}J_{Z}\,\eta^{2}
    .
    \end{equation}
Using the Kramers formula, another numerical coefficient in front  of $\eta^{2}$ was obtained. This is due to the fact that for $\eta\gg1$ the main contribution to the effective radiation is given by transitions to levels with a principal quantum number $n\sim1$, and for these transitions the Kramers formula is not applicable.
\begin{figure}[!t]
  \centering
  \includegraphics[width=\linewidth]{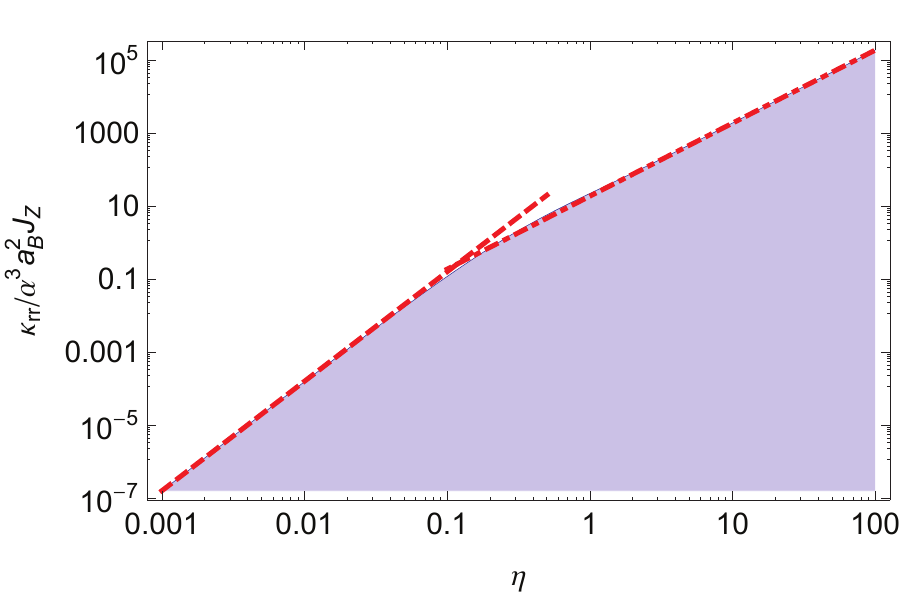}
  \caption{
        Effective radiation $\varkappa _{\text{rr}}$ in units $\alpha^3a_B^2 J_Z$ calculated by exact formula \eqref{4:03} (the edge of the shaded area), approximations \eqref{4:06} for $\eta\ll1 $ (dashed line), and \eqref{4:07} for $\eta\gg1 $ (dash-dotted line).
  }\label{fig:KappaTotal}
\end{figure}
The result of numerical integration in Eq.~\eqref{4:03} is shown in Fig.~\ref{fig:KappaTotal} together with the asymptotics \eqref{4:06} and \eqref{4:07}.

A piecewise-continuous parametrization of the total effective radiation for recombination was proposed by Erdas et al. in Ref.~\cite{Erdas+1993PhysRevA_48_452}. However, their formula seems to contain a typo, since it describes a function with discontinuities. We have obtained a simpler parametrization
    \begin{equation}
    \varkappa_{\text{rr}}
    =
    J_{Z}\,\sigma_{\text{rr}}^{(1)}
    \left(
        \num[round-precision=5]{1.23212}
        +
        \num[round-precision=5]{1.20248} \,\eta^{-2}
    \right)
    ,
    \end{equation}
which has the accuracy better than $1\%$.

\section{Recombination rate coefficient}\label{4}

To obtain the recombination rate coefficient $k_\text{rr}$, Eq.~\eqref{1:02}, in  plasma with a Maxwellian electron distribution function $f_{e}(v)$,  it is necessary to calculate the integral
    \begin{equation*}
    k_{\text{rr}}
    =
    \int_{0}^{\infty }
        \sigma _{\text{rr}}
        \sqrt{\frac{2\varepsilon}{m}}
    \frac{
        2\sqrt{\varepsilon }
        \exp(-\varepsilon /T)
    }{
    	\sqrt{\pi}\,T^{3/2}
    }
    \dif\varepsilon
    .
    \end{equation*}
Passing here to integration over the dimensionless variable $\eta= \sqrt{J_{Z}/\varepsilon}$, we arrive at the expression
    \begin{equation}
    \label{3:02}
    k_{\text{rr}}
    =
    \frac{4}{\sqrt{\pi}}\,
    \alpha c Z
    \left(
        \frac{J_{Z}}{T}
    \right)^{3/2}
    \int_{0}^{\infty }
    \frac{\sigma _{\text{rr}}(\eta)}{\eta^{5}}
    \exp\!\left(-\frac{J_{Z}}{T}\, \eta^{-2}\right)
    \dif\eta
    .
    \end{equation}
The result of integration is a function of the dimensionless parameter $T/J_{Z}$.
Substitution of the small-$\eta$ asymptotics \eqref{2.3:04} for the radiative recombination cross section in Eq.~\eqref{3:02}  leads to divergence of the integral at the upper limit where this asymptotics is not applicable. Consequently, for any value of the parameter $T/J_{Z}$, it is necessary to  take into account correctly the contribution of electrons with low energy $\varepsilon \ll J_{Z}$ (i.e.\ $\eta\gg1$).
If we use  the asymptotics \eqref{2.3:07} in Eq.~\eqref{3:02}, we obtain the integral which can be  taken analytically. At $T/J_{Z} \ll 1$ the result of this integration  reads
    \begin{equation}
    \label{3:03}
    k_{\text{rr}}
    =
    \frac{32\sqrt{\pi}}{3\sqrt{3}}\,
    \alpha^{4} c Z a_{\text{B}}^{2}
    \left(
        \frac{J_{Z}}{T}
    \right)^{1/2}
    \left[
        \ln\left(\frac{J_{Z}}{T}\right)
        +
        \gamma
    \right]
    ,
    \end{equation}
where $\gamma = \num{0.577216}$ is the Euler constant. Previously, this formula was obtained in Ref.~\cite{Bell+1981PartAccel_12_49}.

Seaton in his famous article \cite[Eq.~(36)]{Seaton1959MNRAS_119_81} gave an approximation formula, which in our notations reads
    \begin{multline}
    \label{3:03a}
    k_{\text{rr}}
    =
    \frac{32\sqrt{\pi}}{3\sqrt{3}}\,
    \alpha^{4} c Z a_{\text{B}}^{2}
    \left(
        \frac{J_{Z}}{T}
    \right)^{1/2}
    \\ \times
    \left[
        \ln\left(\frac{J_{Z}}{T}\right)
        +
        \num{0.8576} 
        +
        \num{0.938}\left(\frac{J_{Z}}{T}\right)^{-1/3}
    \right]
    .
    \end{multline}
Seaton's calculations were confirmed in Ref.~\cite{Bell+1981PartAccel_12_49}.
Note that the coefficient in front of the leading term $\ln\left(J_{Z}/T\right)$ in Eqs.~\eqref{3:03} and~\eqref{3:03a} coincide, but the next-to-leading terms differ noticeably. Seaton also derived his result from the Kramers formula, but took into account the deviation of the Gaunt factor from  unity. Recall that the Gaunt factor is a ratio of the result following from  the exact quantum theory and the classical Kramers formula, Eq.~\eqref{2.2:04}. Milstein's formula \eqref{2.3:03} contains this factor de facto, because it was obtained by methods of quantum physics, while in derivation of  Eq.~\eqref{3:03} we actually used the classical Kramers formula. Thus, Seaton's expression \eqref{3:03a} is more accurate than Eq.~\eqref{3:03}. This statement  was checked and confirmed in Ref.~\cite{Tarter1971ApJ_168_313}.
A similar parametrization of the recombination rate coefficient to the excited levels $ k \geq 2 $ was proposed in the article \cite[Eq.~(3.11)]{HummerSeaton1963MNRAS_125_437} by Hummer and Seaton and then confirmed by Brown \& Mathews in their paper \cite[p.~491]{BrownMathews1970ApJ_160_939}.
The latter authors stated that Seaton's result are valid within an error less than $1\%$, but our calculations give slightly different values for the coefficients in Seaton's formula \eqref{3:03a}. We have found that in the interval $10^{-3}<T/J_ {Z}<10 $ the parametrization
    \begin{multline}
    \label{3:03b}
    k_{\text{rr}}
    =
    \frac{32\sqrt{\pi}}{3\sqrt{3}}\,
    \alpha^{4} c Z a_{\text{B}}^{2}
    \left(
        \frac{J_{Z}}{T}
    \right)^{1/2}
    \\ \times
    \left[
        \ln\left(\frac{J_{Z}}{T}\right)
        +
        \num{0.900395}
        +
        \num{0.893211}\left(\frac{J_{Z}}{T}\right)^{-1/3}
    \right]
    \end{multline}
gives a uniform approximation to the exact result with the error less than $1.3\%$, and the maximum error is reached on the edge of the interval at $T/J_{Z}=10 $. Note that Seaton's formula \eqref{3:03a} is given in the well-known NRL formulary on plasma physics \cite[p.~55, Eq.~(13)]{NRL2016}, where its region of applicability is cited as $T/Z^{2}\lesssim 400\,\text{eV}$. However, we found that the error exceeds $50\%$ at the upper edge of the interval and is as large as $9\%$ at $T/J_{Z}=10$.

Using in Eq.~\eqref{3:02} the exact formula \eqref{2.3:03} and fitting the result of numerical integration, we obtained the following asymptotics for  $T/J_{Z}\gg 1$
    \begin{equation}
    \label{3:05}
    k_{\text{rr}}
    =
    \frac{
        \num{138.982}\left(J_{Z}/T\right)^{3/2}
    }{
        1
        +
        \num{5.78687}\left(J_{Z}/T\right)^{\num{0.598497}}
    }
    .
    \end{equation}
The accuracy of this formula is better than $ 2.4\%$ in the interval $1.5<T/J_{Z}<10^{5}$. The temperature dependence $ k_{\text{rr}} \propto T^{-3/2}$ for high temperatures and $k_{\text{rr}} \propto T^{-1/2}$ for low temperatures was predicted in monograph \cite{Vainstein+1973CrossSections(eng)}, but the numerical coefficients in \cite{Vainstein+1973CrossSections(eng)} differ significantly from those found by us.

In multiple editions of his widely cited monograph \cite{Raizer1987(eng), Raizer1992(eng), Raizer2009(eng)} Raizer quotes  the formula $k_{\text{rr}} = 2.7{\times}10^{-13} (T/J_{Z})^{-3/4} [\text{cm}^{4}\text{eV}^{3/4}/\text{sec}]$  referring to the English edition of his other book coauthored with  Zel'dovich \cite{ZeldovichRaizer1966AcademicPress}. We found that his formula in applicable at very narrow interval $0.002 <T/J_{Z}< 6$ where the error varies from $-30\%$ in the mid of the interval to $+30\%$ at its ends.

\begin{figure}[!t]
  \centering
  \includegraphics[width=\linewidth]{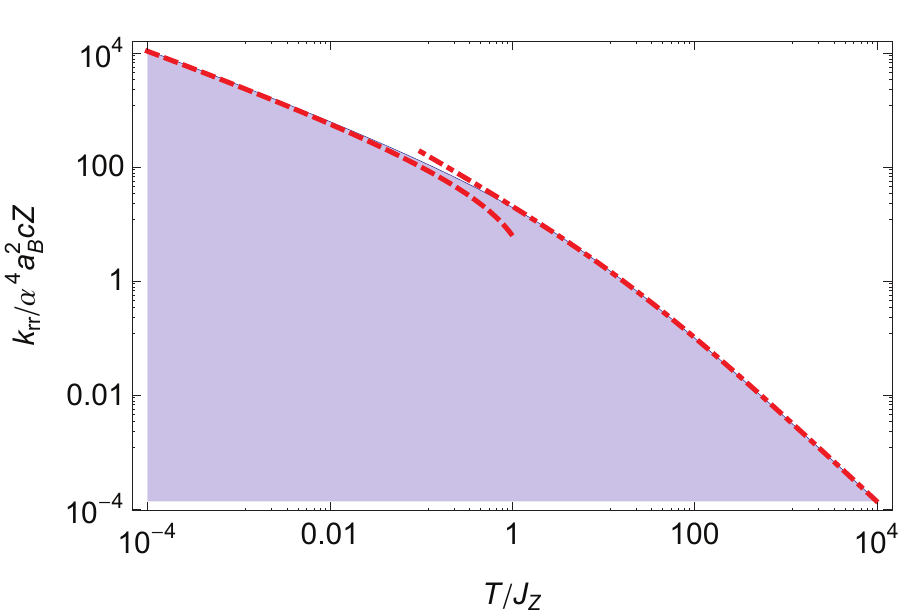}
  \caption{
        Recombination rate coefficient $k_{\text{rr}}$ in units $\alpha^4a_B^2cZ$ calculated by the formula \eqref{3:02} (the edge of dashed area), the approximation \eqref{3:03b} for $T/J_ {Z} \ll 1$ (dashed line), and \eqref{3:05} for $T/J_ {Z}\gg1 $ (dash-dotted line).
  }\label{fig:kTotal}
\end{figure}
\begin{figure}[ht]
  \centering
  \medskip
  \includegraphics[width=\linewidth]{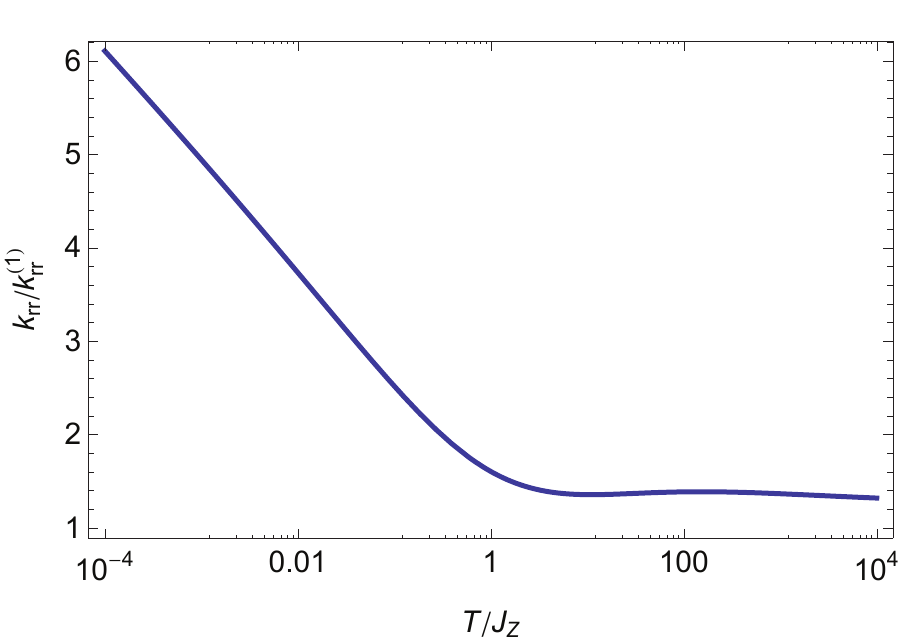}
  \caption{
   Dependence of  $ k_{\text{rr}}/k_{\text{rr}}^{(1)}$ on  $T/J_{Z}$.
  }\label{fig:kTotalTok1}
\end{figure}

The result of numerical integration over a wide range of values of the ratio $T/J_{Z}$ is shown in Fig.~\ref{fig:kTotal}, where the asymptotics \eqref{3:03b} and \eqref{3:05} are also plotted.
Combining these asymptotics we derived the following uniform parametrization
    \begin{equation}
    k_{\text{rr}}
    =
    \frac{
        \num{8.41413} \left[
            \ln\left(
                1+J_{Z}/T
            \right)
            +
            \num{3.49906}
        \right]
        \alpha ^{4} c Z a_{\mathrm{B}}^{2}
    }{
        (
            T/J_{Z})^{1/2}
            + \num{0.651673}\,(T/J_{Z}
        )
        + \num{0.213789}\,(T/J_{Z})^{3/2}
    }.
    \end{equation}
In the range $10^{-4} < T/J_{Z} < 10^{4}$, its accuracy is $3\%$. For  the first bound state,  a uniform parametrization of the recombination rate  coefficient reads
    \begin{equation}
    k_{\text{rr}}^{(1)}
    =
    \frac{
        \num[round-precision=4]{17.405864073215675}\,
        \alpha ^{4} c Z a_{\mathrm{B}}^{2}
    }{
        \left(T/J_{Z}\right)^{1/2}
        +\num{0.35931257542154577}\left(T/J_{Z}\right)^{7/6}
        +\num{0.14714777052064387}\left(T/J_{Z}\right)^{3/2}
    }
    .
    \end{equation}
Its accuracy for $T/J_{Z}<10^{4} $ is better than $3\%$. The  ratio $ k_{\text{rr}}/k_{\text{rr}}^{(1)}$ is shown  in Fig.~\ref{fig:kTotalTok1}. Note that the limit of $k_{rr}^{(1)}=17.5\,\alpha ^{4}cZa_{\mathrm{B}}^{2}$ at $T\to0$ was calculated in Ref.~\cite[\S24, problem 1]{LLX1981(eng)}.

Finally, we  mention the work \cite[Eq.~(4)]{Verner+1996AJSS_103_467} where  a four-parametric formula is proposed for the recombination rate coefficients for various ions, which, according to the authors, has an accuracy better than $3\%$ in the interval $3 {\times} 10^{-5} <T/J_{Z} < 10^{5}$. The same formula is reproduced by the authors of the review article \cite[Eq.~(6)]{Mazzotta+1998AASS_133_403}. Our calculations has shown that at least in the case of hydrogen ions, this parametrization overestimates the  recombination rate coefficient  by many times.

\medskip

\section{The  emission coefficient}\label{5}

A power of  radiation in the recombination process (i.e. the energy emitted per unit volume per unit time) is given  by the formula
    \begin{equation}
    \label{5:01}
    P_{\text{rr}}
    =
    n_{i} n_{e}
    q_{\text{rr}}
    \end{equation}
where
    \begin{equation*}
    q_{\text{rr}}
    =
    \int_{0}^{\infty}
        \varkappa_{\text{rr}}
        \sqrt{\frac{2\varepsilon }{m}}
    \frac{
        2\sqrt{\varepsilon }\exp(-\varepsilon /T)
    }{
        \sqrt{\pi}\,T^{3/2}
    }
       \dif\varepsilon
    \end{equation*}
is  the  emission coefficient \eqref{1:02}. It can also be written in the form
    \begin{equation}
    \label{5:02}
    q_{\text{rr}}
    =
    \frac{4}{\sqrt{\pi}}\,
    \alpha c Z
    \left(
        \frac{J_{Z}}{T}
    \right)^{3/2}
    \int_{0}^{\infty }
    \frac{\varkappa_{\text{rr}}(\eta)}{\eta^{5}}
    \exp\!\left(-\frac{J_{Z}}{T}\, \eta^{-2}\right)
    \dif\eta
    .
    \end{equation}
Substituting in the last expression the asymptotics \eqref {4:06} for $\eta\ll1$, we arrive at the asymptotic expression
    \begin{equation}
    \label{5:03}
    q_{\text{rr}}
    =
    \frac{256}{3} \pi  \zeta (3)
    \,
    \alpha ^{4}  c Z J_{Z} a_{\text{B}}^{2}
    \frac{J_{Z}}{T}
    \end{equation}
for the  emission coefficient at $ T\gg J_{Z} $. We did not find an analogue of this formula in the literature. This may be due  to the fact that a power of recombination radiation at  high temperatures is significantly smaller than the power of bremsstrahlung.

In the opposite case $ T \ll J_ {Z} $, the dependence of $q_{\text{rr}}$ on temperature is different, namely:
    \begin{equation}
    \label{5:04}
    q_{\text{rr}}
    =
    \num{21.4874}
    \,
    \alpha ^{4}c Z J_{Z} a_{\text{B}}^{2}
    \sqrt{\frac{J_{Z}}{T}}
    .
    \end{equation}
Our result is approximately 18\% less than that of Eq.~(25) in the review article of Kogan \&  Lisitsa \cite{KoganLicitsa1983VINITI_4_194(eng)}. That result was  derived in the Kramers approximation; the authors also refer to numerical calculations in early paper by Kogan \cite[Eq.~(9)]{Kogan1958FizPlazUTS3_99(eng)}. Our formula \eqref{5:04} should also be compared with Eq.~(33) on page 58 of the reference book \cite{NRL2016} for which conditions of applicability are not specified. Although the temperature and $ Z $ dependence   are the same,  our coefficient is $ 1.8 $ times larger.

Seaton in the above mentioned  paper \cite[Eq.~(52)]{Seaton1959MNRAS_119_81} derived an asymptotic formula for what he called the total kinetic energy loss due to radiative recombination. His equation resembles Eq.~\eqref{3:03a} for the recombination rate coefficient at low electron energies $\varepsilon$. Unfortunately, a direct comparison of this Seaton's result with our calculation is not possible as the total emission coefficient of radiative recombination is larger by definition (emitted photon brings out the energy  $\varepsilon$ of free electron minus the energy  $-J_{Z}/n^{2}$ of the electron in a recombined state).

\begin{figure}
  \centering
  \includegraphics[width=\linewidth]{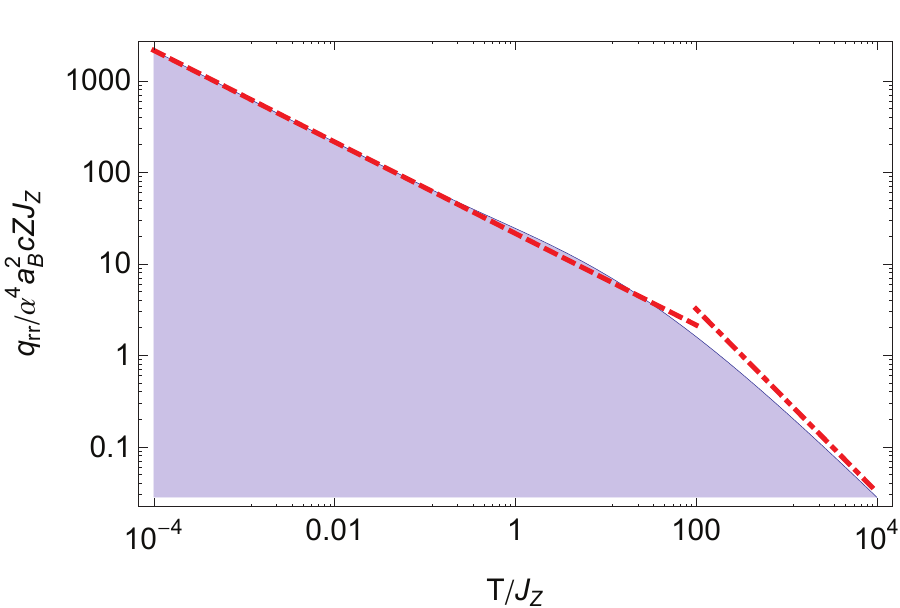}
  \caption{
        Emission coefficient $q_{\text{rr}}$ in units $\alpha^4a_B^2cZJ_Z$ (the border of the shaded area), the approximation \eqref{5:04} for $T\ll J_{Z} $ (dashed line) and \eqref{5:03} for $ T\gg J_ {Z} $ (dash-dotted line).
  }\label{fig:PowerTotal}
\end{figure}

The result of evaluation of Eq.~\eqref{5:02} is shown in Fig.~\ref{fig:PowerTotal}. It is seen  that the asymptotics \eqref{5:03} works well enough  only for an extremely large ratio $T/J_{Z}\gtrsim 10^{3}$.  We suggest a uniform parametrizations for $q_{\text{rr}}$,
    \begin{equation}
    \label{5:06}
    q_{\text{rr}}
    =
    \frac{
        \left[
            \num{20.9293}
            +
            \num{18.6447}\,(T/J_{Z})^{1/2}
        \right]
        \alpha^{4}c Z J_{Z}
    }{
        (T/J_{Z})^{1/2}
        +\num{0.561279}\, (T/J_{Z})
        +\num{0.0612936 }\, (T/J_{Z})^{3/2}
    }\,,
    \end{equation}
and  for transition to the first level
    \begin{equation}
    \label{5:07}
    q_{\text{rr}}^{(1)}
    =
    \frac{
        \left[
            \num{17.0462}
            +
            \num{14.1953 }\,(T/J_{Z})^{1/2}
        \right]
        \alpha^{4}c Z J_{Z}
    }{
        (T/J_{Z})^{1/2}
        +\num{0.515988 }\, (T/J_{Z})
        +\num{0.0560782 }\, (T/J_{Z})^{3/2}
    }
    .
    \end{equation}
In the range of $10^{-5}<T/J_{Z}<10^{5}$, the accuracy of both parametrizations is better than $3\%$.
In Ref.~\cite{ErdasQuarati1994ZPhysD_31_161}, Erdas \& Quarati presented results of tabulation of the emission coefficient in a wider interval  $10^{-6}<T/J_Z<10^{8}$, but we concluded that these authors overestimated $q_{\text{rr}}$ by 5\% to 70\% in the range $3{\times}10^{-4} < T/J_{Z} < 1$.

\begin{figure}[!tb]
  \centering
  \medskip
  \includegraphics[width=\linewidth]{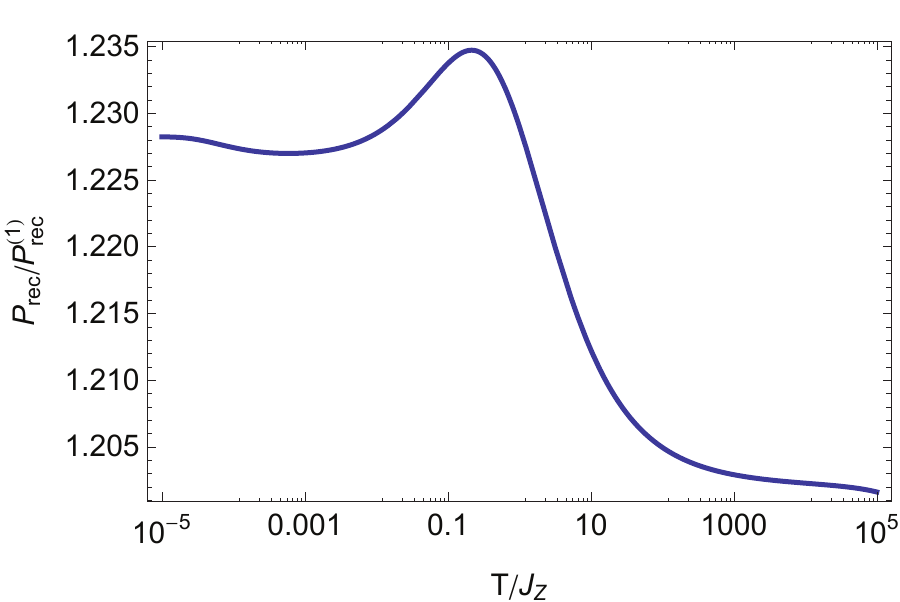}
  \caption{
        Ratio $q_{\text{rr}}/q_{\text{rr}}^{(1)}$ as a function of $T/J_Z$.
  }\label{fig:PTotalToP1}
\end{figure}
The  ratio $q_{\text{rr}}/q_{\text{rr}}^{(1)}$ is shown in Fig.~\ref{fig:PTotalToP1}. It is seen  that  transition of an electron to the first level gives  the main contribution to $q_{\text{rr}}$  for any temperatures. Transition to excited levels gives the contribution to  $q_{\text{rr}}$ from $20$\% to $23.5$\%. In contrast, the contribution of the excited levels to the recombination rate coefficient $k_{\text{rr}}$ at $T\lesssim J_{Z}$ is essentially greater than the contribution of the ground level, see Fig.~\ref{fig:kTotalTok1}.

\medskip
\section{Summary}\label{9}
Our work brings a line under the long-term research of the process of radiation recombination. Using the exact convenient expressions for the cross section and the effective radiation, we have formulated a status of numerous approximate  results concerning the  process under discussion. For the total radiative recombination cross section, the total recombination effective radiation, the recombination rate coefficient and the emission coefficient, we have also suggested new uniform interpolation  formulas having high accuracy in a wide range of electron energies and plasma temperatures.

\begin{acknowledgements}
    The work was supported by the Russian Science Foundation (Project No.~14-50-00080).

\end{acknowledgements}


%

\end{document}